\documentclass[
    ,final            
  ]
  {aipproc}

\layoutstyle{8x11double}

\newcommand\Lunit {ergs s$^{-1}$}

\newcommand\mzon   {M$_{\odot}$}

\def\ga{\mathrel{\mathchoice {\vcenter{\offinterlineskip\halign{\hfil
$\displaystyle##$\hfil\cr>\cr\noalign{\vskip1.5pt}\sim\cr}}}
{\vcenter{\offinterlineskip\halign{\hfil$\textstyle##$\hfil\cr>\cr
\noalign{\vskip1.0pt}\sim\cr}}}
{\vcenter{\offinterlineskip\halign{\hfil$\scriptstyle##$\hfil\cr>\cr
\noalign{\vskip0.5pt}\sim\cr}}}
{\vcenter{\offinterlineskip\halign{\hfil$\scriptscriptstyle##$\hfil
\cr>\cr\noalign{\vskip0.5pt}\sim\cr}}}}}

\def\la{\mathrel{\mathchoice {\vcenter{\offinterlineskip\halign{\hfil
$\displaystyle##$\hfil\cr<\cr\noalign{\vskip1.5pt}\sim\cr}}}
{\vcenter{\offinterlineskip\halign{\hfil$\textstyle##$\hfil\cr<\cr
\noalign{\vskip1.0pt}\sim\cr}}}
{\vcenter{\offinterlineskip\halign{\hfil$\scriptstyle##$\hfil\cr<\cr
\noalign{\vskip0.5pt}\sim\cr}}}
{\vcenter{\offinterlineskip\halign{\hfil$\scriptscriptstyle##$\hfil
\cr<\cr\noalign{\vskip0.5pt}\sim\cr}}}}}

\def\degr{\hbox{$^\circ$}}

\begin{document}

\title{Observations of millisecond X-ray pulsars}

\author{Rudy Wijnands}{
  address={Astronomical Institute 'Anton Pannekoek', University of
  Amsterdam, 1098 SJ Amsterdam, The Netherlands} }

\begin{abstract}
I present an observational review of the five accretion-driven
millisecond X-ray pulsars currently known, focusing on the results
obtained with the {\it Rossi X-ray Timing Explorer} ({\it RXTE})
satellite.  A prominent place in this review is given to the first
such system discovered, SAX J1808.4--3658. Currently four outbursts
have been detected from this source, three of which have been studied
using {\it RXTE}. This makes this source the best studied example of
all accretion-driven millisecond pulsars. Its October 2002 outburst is
of particular interest because of the discovery of kilohertz
quasi-periodic oscillations and burst oscillations during its
thermonuclear X-ray bursts.  The other four accreting millisecond
pulsars were discovered within the last two years and only limited
results have been published so far. A more extended review can be
found at http://zon.wins.uva.nl/$\sim$rudy/admxp/

\end{abstract}

\maketitle


\section{Introduction}

Ordinary pulsars are highly-magnetized ($B\sim 10^{12}$ G), rapidly
rotating ($P\sim 10$ ms) neutron stars which spin down on timescales
of 10 to 100 million years due to magnetic dipole radiation. By
contrast, millisecond ($P<10$ ms) radio pulsars have ages of billions
of years and weak ($B\sim 10^{8-9}$ G) surface magnetic fields.  Since
many of these millisecond radio pulsars are in binary systems, it has
long been suspected (see, e.g., \cite{bvdh1991} for an extended
review) that the neutron stars were spun up by mass transfer from a
stellar companion in a low-mass X-ray binary (LMXB). If this scenario
is correct, LMXBs should harbor a fast spinning neutron star which
might be visible as an accretion-driven millisecond X-ray pulsar.
However, before the launch of the {\itshape Rossi X-ray Timing
Explorer} ({\itshape RXTE}), all searches for coherent millisecond
X-ray pulsations in LMXBs failed to yield a detection \cite[and
references therein]{v1994}. Only after {\it RXTE} was launched did we
obtain conclusive evidence that at least some LMXBs indeed harbor
weakly magnetic neutron stars with millisecond spin periods.  In April
1998 the first accreting millisecond X-ray pulsar was discovered
\cite{wvdk1998} followed by the discovery of four additional systems
during the last two years
\cite{markwardt2002,galloway2002,markwardt2003_1807,markwardt2003_1814}.

\section{SAX J1808.4--3658 }

\subsection{The September 1996 outburst}

In September 1996, a previously unknown X-ray transient (designated
SAX J1808.4--3658) was detected with the Wide Field Cameras aboard the
{\it BeppoSAX} satellite \cite{intzand1998}. Three thermonuclear X-ray
bursts were detected, demonstrating that the system harbors a neutron
star. Using those type-I X-ray bursts the distance toward the source
was estimate to be 2.5 kpc \cite{intzand1998,intzand2001}, resulting
in a maximum outburst luminosity of $\sim10^{36}$ \Lunit.  The
outburst lasted for about three weeks, after which the source was
presumed to have returned to quiescence.  However, recently it was
found
\cite{revnivtsev2003} that the source could still be detected on
October 29, 1996, (using data obtained with the proportional counter
array [PCA] aboard {\itshape RXTE} during slew maneuvers of the
satellite) with a luminosity of approximately a tenth of the outburst
peak luminosity.  This demonstrates that six weeks after the main
outburst the source was still active which might indicate that the
behavior of the source at the end of the 1996 outburst was similar to
what has been seen during the 2000 and 2002 outbursts (see below).

After SAX J1808.4--3658 was found to harbor a millisecond pulsar
\cite{wvdk1998}, the X-ray bursts  seen with {\it BeppoSAX}
were scrutinized for potential burst oscillations
\cite{intzand2001}. A marginal detection of a 401 Hz oscillation was
made in the third burst suggesting that the burst oscillations
observed in the other, non-pulsating, neutron-star LMXBs indeed occur
at their neutron-star spin frequencies. This result has been confirmed
by the recent detection of burst oscillations during the 2002 outburst
of SAX J1808.4--3658 (see below;
\cite{chakrabarty2003}).

\subsection{The April 1998 outburst}

\begin{figure}
\includegraphics[width=7.5cm]{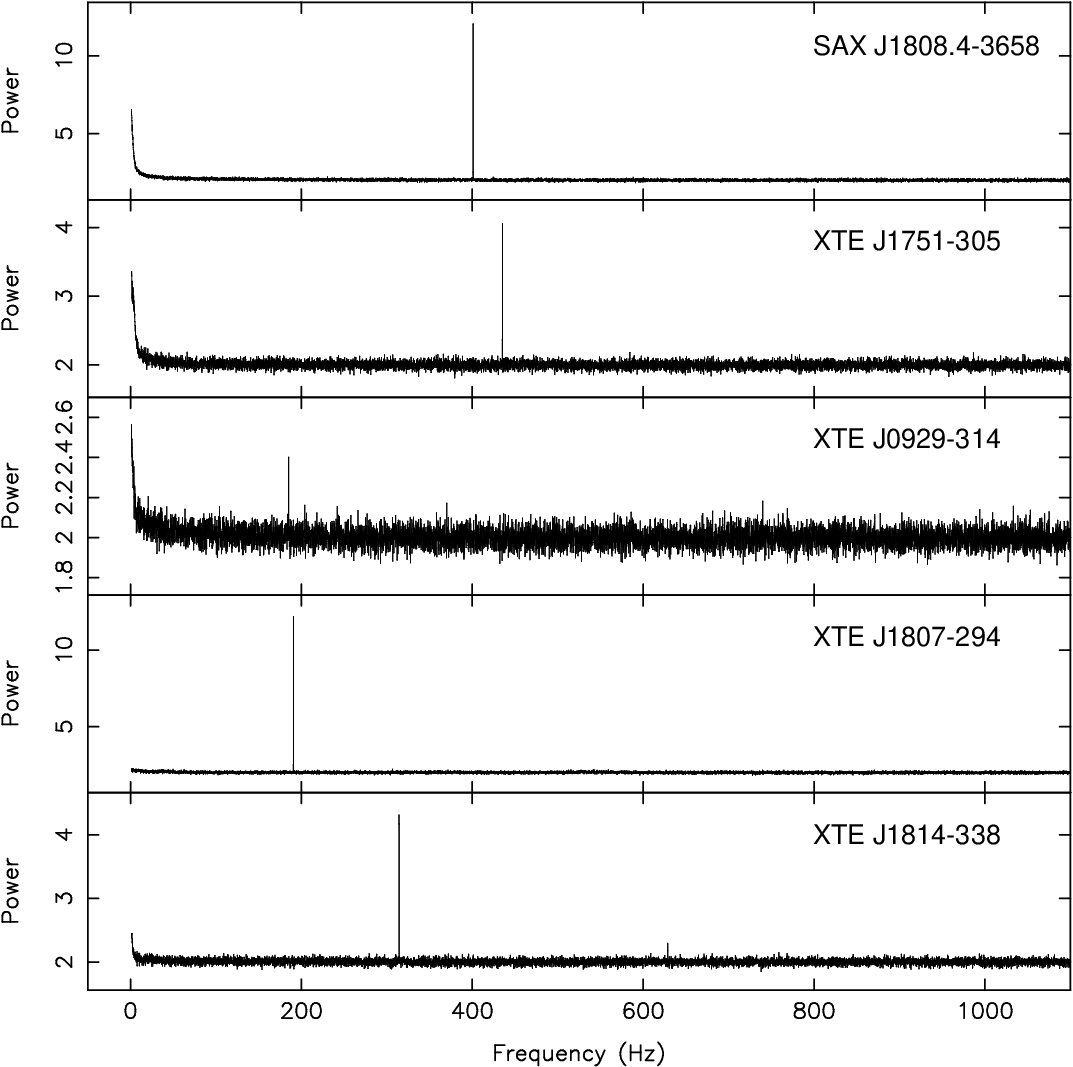}
\caption{
Examples of the power spectra obtained for the five millisecond
pulsars showing the pulsar spikes.
\label{fig:pulsations}}
\end{figure}

On April 9, 1998, SAX J1808.4--3658 was found to be active again
\cite{marshall1998} and a public TOO observation campaign on the
source was started using the {\it RXTE}/PCA.  From those observations,
it was discovered that coherent 401 Hz pulsations were present in the
persistent X-ray flux of the source (\cite{wvdk1998};
Fig.~\ref{fig:pulsations}), making this source the first accreting
millisecond X-ray pulsar discovered. An analysis of the coherent
timing properties of the source showed that the neutron star is in a
$\sim$2-hr binary system with a very low mass companion
\cite{cm1998}. The limited amount of {\it RXTE}/PCA data obtained
during this outburst yielded only an upper limit of $<7\times
10^{-13}$ Hz s$^{-1}$ on the pulse-frequency derivative \cite{cm1998}.

The source flux during the 1998 outburst first showed a steady decline
in X-ray flux, which accelerated after $\sim$2 weeks
(\cite{gilfanov1998,cui1998}; Fig.~\ref{fig:1808_lc_2002}). This
behavior has been attributed to the fact that the source might have
entered the 'propeller regime' in which the accretion of matter is
centrifugally inhibited
\cite{gilfanov1998}. However, the fact that after the onset of the
steep decline the pulsations could still be detected \cite{cui1998}
makes this interpretation doubtful. A week after the onset of this
steep decline, the flux leveled off
\cite{cui1998,wang2001}, but  the X-ray flux behavior of the source at the
end of the outburst remains unclear, since no further {\itshape
RXTE}/PCA observations were made. It is possible that the source might
have displayed a similar long-term episode of low-luminosity activity
as seen at the end of the 2000 and 2002 outbursts (see below and
Fig.~\ref{fig:1808_lc_2002}).

Studies of the X-ray spectrum (\cite{gilfanov1998,hs1998}; see also
\cite{gierlinski2002, poutanengierlinski2003}) and the aperiodic rapid
X-ray variability (\cite{wvdk1998_bbn,vanstraaten2004}; see also
Fig.~\ref{fig:broad-band}) showed an object that, apart from its
pulsations, is remarkably similar to LMXBs with comparable
luminosities (the atoll sources). There is apparent modulation of the
X-ray intensity at the orbital period, with a broad minimum when the
pulsar is behind the companion \cite{cm1998,hs1998}.  Cui et
al.~\cite{cui1998} and Ford \cite{ford1999} reported on the harmonic
content, the energy dependency, and the soft phase lag of the
pulsations. The main result of those studies is that the low-energy
pulsations lag the high-energy ones by as much as $\sim$$200 \mu$s
($\sim$8\% of the pulsation period; see
\cite{cui1998,ford1999,poutanengierlinski2003} for 
explanations).

SAX J1808.4--3658 was also detected in the optical, IR, and in radio
bands. Its optical/IR counterpart (later named V4580 Sgr;
\cite{kazarovets2000}) was discovered by Roche et al. \cite{roche1998}
and subsequently confirmed by Giles et al. \cite{giles1998}. A
detailed study of the optical behavior during this outburst can be
found in Giles et al. \cite{giles1999} and Wang et
al. \cite{wang2001}. Giles et al. \cite{giles1999} also reported that
the optical flux was modulated at the 2-hr orbital period. A model of
the X-ray and optical emission from the system using an X-ray-heated
accretion disk model, gave a best fit values of the $A_v$ of 0.68 and
the inclination of $\cos i = 0.65$ \cite{wang2001}, resulting in a
mass of the companion of 0.05--0.10 \mzon. Some of the IR fluxes were
too high to be consistent with emission from the disk or the companion
star, even when considering X-ray heating. This IR excess might be due
to synchrotron processes, possibly related to an outflow or ejection
of matter \cite{wang2001}. Such an event was also suggested by the
discovery of the radio counterpart
\cite{gaensler1999}. The source was detected with a 4.8 GHz flux of
$\sim$0.8 mJy on April 27, 1998, but it was not detected at other
epochs.

\subsection{The  January 2000 outburst }

\begin{figure}
\includegraphics[angle=-90,width=7.5cm]{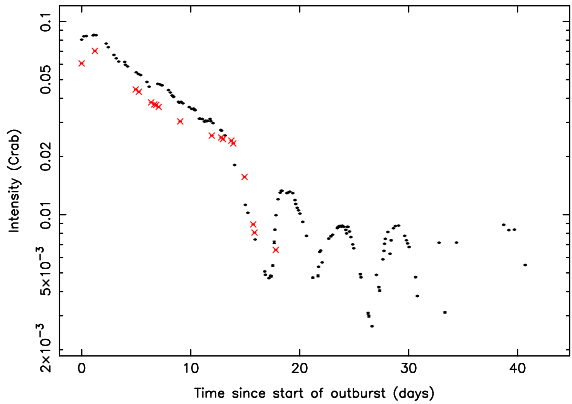}
\caption{
The {\it RXTE}/PCA light curves of SAX J1808.4--3658 during its 1998
(red crosses) and 2002 outbursts (black dots). The data were taken from van
Straaten et al. \cite{vanstraaten2004} \label{fig:1808_lc_2002}}
\end{figure}

On January 21, 2000, SAX J1808.4--3658 was again detected
\cite{wijnands2001} with the {\it RXTE}/PCA but this time at a flux level of
$\sim$10--15 mCrab (2--10 keV). This is only about a tenth of the
fluxes observed during the peak of the previous
outbursts. Furthermore, it was found to exhibit low-level activity for
months \cite{wijnands2001}. Due to solar constraints the source could
not be observed before January 21 but it is likely that a true
outburst occurred before that date and only the end stages of this
outburst could be observed. This is supported by the very similar
behavior of the source observed near the end of its 2002 outbursts
(Fig.~\ref{fig:1808_lc_2002}).

During the 2000 outburst, the source was observed (using {\itshape
RXTE}/PCA) at luminosities of $\sim$$10^{35}$ ergs s$^{-1}$ on some
occasions, but on others (a few days earlier or later) the source had
luminosities of only $\sim$$10^{32}$ ergs s$^{-1}$ (as seen during
{\itshape BeppoSAX} and {\itshape XMM-Newton} observations
\cite{wijnands2002,wijnands2003}). This demonstrates that SAX
J1808.4--3658 exhibited extreme luminosity variations by a factor of
$>1000$ on timescales of days.  During the episodes of low-level
activity in the {\it RXTE}/PCA observations it was also found that on
several occasions SAX J1808.4--3658 exhibited strong (up to 100\% rms
amplitude) and violent flaring behavior with a frequency of $\sim$1 Hz
(\cite{vdk2000_1808,wijnands2001_flaring}; Fig. 1 in
\cite{wijnands2003_review}). During this episode of low-level
activity, the pulsations were also detected but the limited amount of
observing time and the low source flux did not allow for an
independent determination of the orbital parameters and the
pulse-frequency derivative.

In optical, the source was also detected, albeit at a lower
brightness than during the 1998 outburst \cite{wh2000}, consistent
with the lower observed X-ray fluxes. The source was frequently
observed during this outburst and the results were presented by Wachter et
al. (\cite{wachter2000}; see also the discussion by Wijnands
\cite{wijnands2003_review}).

\subsection{The October 2002 outburst}

In October 2002, the fourth outburst of SAX J1808.4--3658 was detected
\cite{markwardt2002_1808}, immediately launching an  extensive {\itshape
RXTE}/PCA observing campaign. The main results are summarized below.

\subsubsection{The X-ray light curve}

The light curve for this outburst is shown in
Figures~\ref{fig:1808_lc_2002} and \ref{fig:lightcurves}. During the
first few weeks, the source decayed steadily, until the rate of
decline suddenly increased, similar to what was observed during the
1998 outburst (in Fig.~\ref{fig:1808_lc_2002} the 1998 outburst light
curves is also plotted for comparison). During both the 1998 and 2002
outbursts, the moment of acceleration of the decline happened at about
two weeks after the peak of the outburst.  Approximately five days
later the X-ray fluxes rapidly increased again until it reached a peak
of about a tenth of the outburst maximum. After that SAX J1808.4--3658
entered a state in which its flux fluctuated rapidly on time scales of
days to hours, very similar to the low-level activity seen during its
2000 outburst. The 2002 outburst light curve is the most detailed one
seen for SAX J1808.4--3658 and it exhibits all features seen during
its previous outbursts (the initial decline, the increase in the
decline rate, the long-term activity), demonstrating that this is
typical behavior for this source.

\subsubsection{The kHz QPOs}

\begin{figure}[t]
\includegraphics[width=7.5cm]{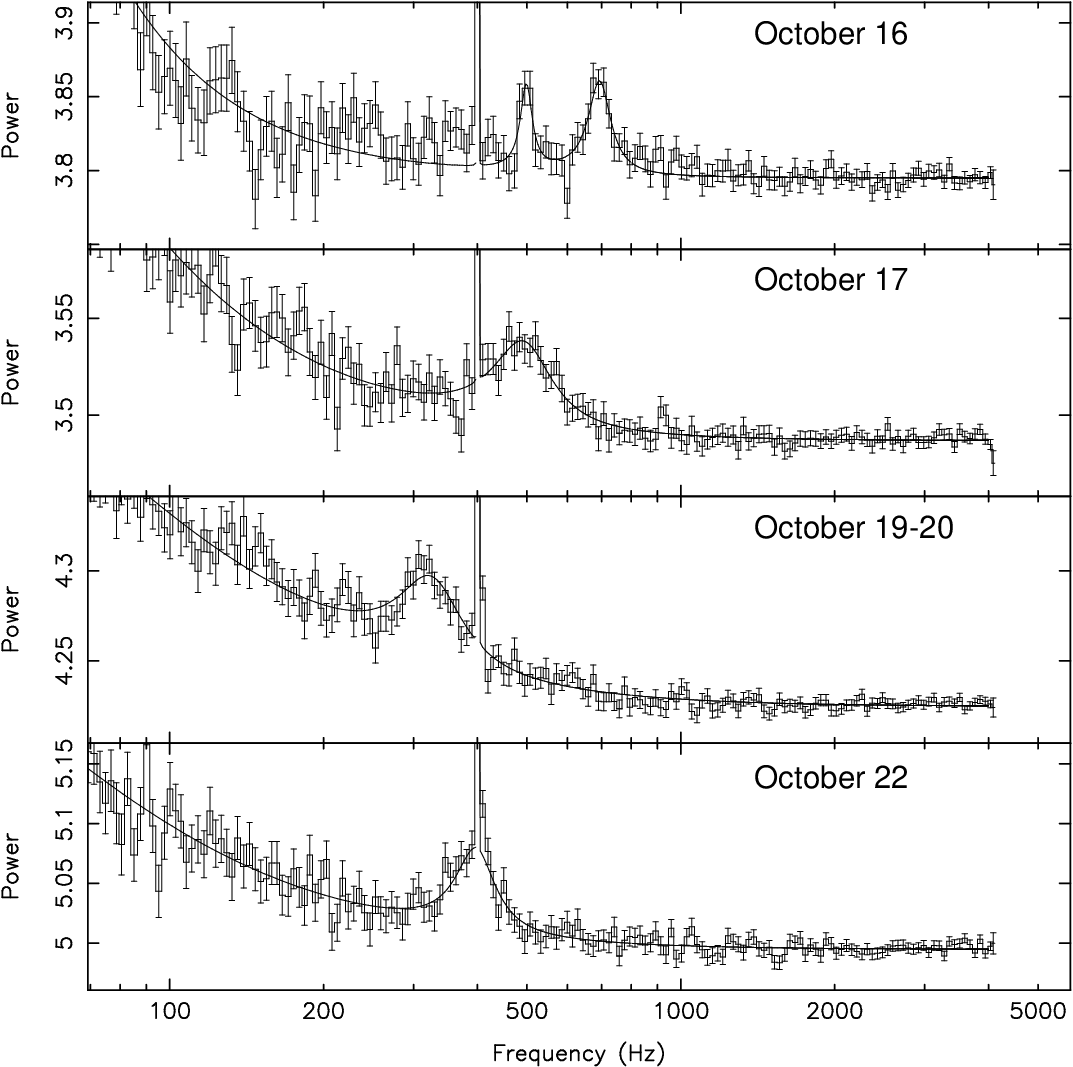}
\caption{
The power spectra obtained for SAX J1808.4--3658 during its 2002
outburst showing the kHz QPOs. 
\cite{wijnands2003_nature} \label{fig:1808_2002_kHzQPOs}}
\end{figure}

Wijnands et al. \cite{wijnands2003_nature} reported the discovery of
two simultaneous kHz QPOs during the peak of the 2002 outburst, with
frequencies of $\sim$700 and $\sim$500 Hz
(Fig.~\ref{fig:1808_2002_kHzQPOs} top), making this the first
detection of twin kHz QPOs in a source with a known spin-frequency.
The frequency separation between the two peaks is $\sim$200 Hz. This
is significantly below the 401 Hz expected in the beat-frequency
models proposed to explain the kHz QPOs and, therefore, those models
are falsified by the detection of the kHz QPOs in SAX
J1808.4--3658. However, the fact that the peak separation is
consistent with half the spin frequency suggests that the kHz QPOs are
indeed connected to the neutron-star spin frequency, albeit in a way
not predicted by any existing model. The lower-frequency kHz QPO was
only seen on October 16, 2002 (i.e., during the peak of the outburst)
but the higher-frequency kHz QPO could be traced throughout the main
part of the outburst (\cite{wijnands2003_nature};
Fig.~\ref{fig:1808_2002_kHzQPOs}).  Besides the twin kHz QPOs, a third
kHz QPO was observed with frequencies ($\sim$410~Hz) just exceeding
that of the pulse \cite{wijnands2003_nature}. The nature of this QPO
is currently still unclear (see \cite{wijnands2003_nature} for a
discussion).

Wijnands et al. \cite{wijnands2003_nature} pointed out that there
appear to exist two classes of neutron-star LMXBS: the 'fast' and the
'slow' rotators. The fast rotators have spin frequencies $\ga$400 Hz
and the peak separation between the kHz QPOs is approximately equal to
half the spin frequency. In contrast, the slow rotators have spin
frequencies below $\la$400 Hz and a peak separation roughly equal to
the spin frequency. The kHz QPO results obtained for SAX J1808.4--3658
have already spurred new theoretical investigations into the nature of
the kHz QPOs \cite{kluzniak2003,lm2003}.

\subsubsection{X-ray bursts and burst oscillations}

During the first five days of the 2002 outburst, four type-I X-ray
bursts were observed, all of which exhibited burst oscillations
\cite{chakrabarty2003}. The frequency in the burst tails was constant
and identical to the spin frequency, while the oscillation in the
burst rise showed evidence of a rapid frequency drift of up to 5
Hz. No oscillations were seen during the peak of the bursts.  This
behavior is similar to the burst oscillations seen in other,
non-pulsating neutron-star LMXBs, demonstrating that
burst-oscillations indeed occur at the neutron-star spin frequency in
all sources. The spin frequency is now known for 16 LMXBs with the
highest spin frequency being 619 Hz. Chakrabarty et
al. \cite{chakrabarty2003} used the sample of burst-oscillation
sources to demonstrate that neutron stars in LMXBs spin well below
their break-up frequency. This could suggest that the neutron stars
are limited in their spin frequencies, possibly due to gravitational
radiation.

\subsubsection{The pulsations}

Pulsations were detected at all flux levels with an amplitude of
3\%--10\%.  The pulsar was spinning down at a constant rate (mean
spin-down rate of $2\times 10^{-13}$ Hz s$^{-1}$;
\cite{chakrabarty2003}), despite a large dynamic range in X-ray flux.
The magnitude of the pulse-frequency derivative exceeds the maximum
value expected from accretion torques by a factor of 5. The timing
history also contains a small glitch with a very rapid recovery time
scale.  There was no evidence for a 200.5 Hz subharmonic in the data
(upper limit of 0.38\% of the signal at 401
Hz;\cite{wijnands2003_nature}) confirming the interpretation of 401 Hz
as the pulsar spin frequency. A detailed analysis will be presented
elsewhere \cite{morgan2004}.

\subsubsection{The low-frequency QPOs}

During the peak of the outburst and subsequent decay, broad-noise and
QPOs with frequencies between 10 and 80 Hz were detected in the power
spectra (\cite{wijnands2003_review}; Fig.~\ref{fig:broad-band}).
Similar phenomena have been observed in other non-pulsating systems
and are likely related to the noise components seen in SAX
J1808.4--3658. Van Straaten et
al. \cite{vanstraaten2003,vanstraaten2004} have studied the broad-band
power spectra (including the noise components, the low-frequency QPOs,
and the kHz QPOs) of SAX J1808.4--3658 in detail as well as the
frequency correlations between the different power-spectral
components.  Interestingly, using those frequency correlations, van
Straaten et al. \cite{vanstraaten2003} suggested that the
higher-frequency kHz QPO could also be identified during the 1998
outburst but at the lowest frequencies found so far in any kHz QPO
source (down to $\sim$150 Hz). Previous work
\cite{wvdk1998_bbn} on the aperiodic timing features of SAX
J1808.4--3658 during its 1998 outburst already found these features
but they could not be identified as the higher-frequency kHz QPO due
to their low frequency and broad character.

Van Straaten et al. \cite{vanstraaten2003,vanstraaten2004} also
compared the results for SAX J1808.4--3658 with those obtained for
other, non-pulsating neutron-star LMXBs and found that the frequency
correlations in SAX J1808.4--3658 are similar to those seen in the
other non-pulsating sources, but that they show a shift in the
frequency of the kHz QPOs. It is unclear what the physical
mechanism(s) is behind this difference between sources
\cite{vanstraaten2003,vanstraaten2004}.
 
As can be seen from Figure~\ref{fig:1808_lc_2002}, during both the
1998 and 2002 outbursts of SAX J1808.4--3658, the source exhibited
similar X-ray fluxes. However, at similar flux levels, the
characteristic frequencies observed during the 1998 outburst are much
lower (factor 10) than during the 2002 outburst
(\cite{vanstraaten2004}; see also Fig.~\ref{fig:broad-band}). Again it
is unclear what causes this huge difference between the two outbursts
but it might be related to the 'parallel track' phenomena observed for
the kHz QPOs (e.g., \cite{vdk2000}).

\subsubsection{The violent 1 Hz flaring}

\begin{figure}
\includegraphics[width=6.5cm]{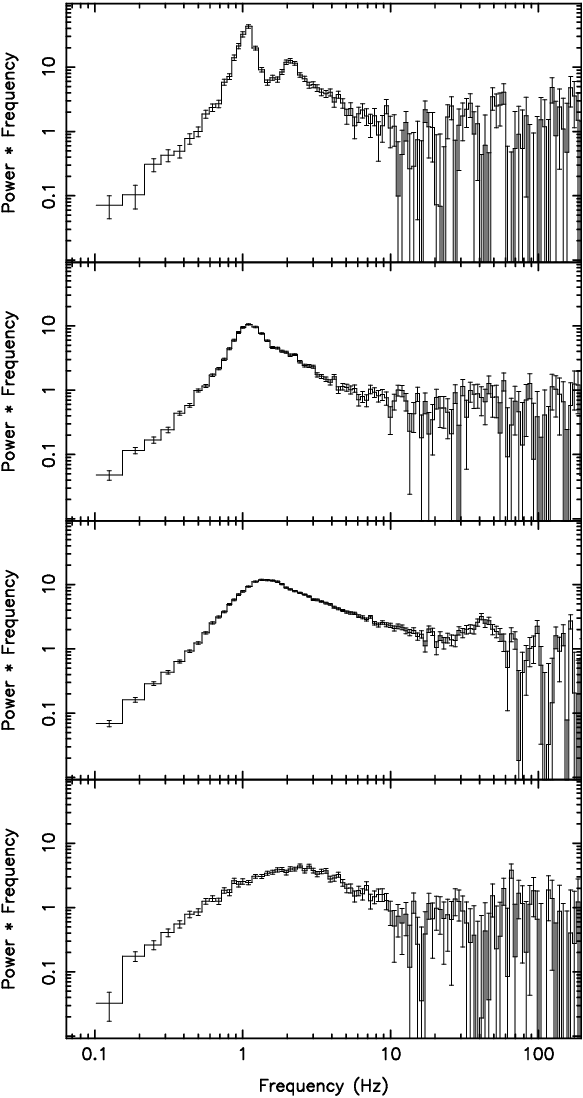}
\caption{
The 1 Hz flaring (represented via power spectra) as observed
during the 2002 outburst of SAX J1808.4-3658.
\label{fig:1808_2002_1Hzflares}}
\end{figure}

Violent flaring was observed on many occasions at a $\sim$1 Hz
repetition frequency during the late stages of the 2002 outburst
(Fig.~\ref{fig:1808_2002_1Hzflares}), similar to what had been
observed during the 2000 outburst. This proves that also this violent
flaring is a recurrent phenomenon and can likely be observed every
time the source is in this prolonged low-level activity
state. Preliminary results presented in
Figure~\ref{fig:1808_2002_1Hzflares} show examples of power spectra
obtained during the end stages of the 2002 outburst. During certain
observations the 1 Hz QPO is rather narrow and its first overtone can
be seen clearly (Fig.~\ref{fig:1808_2002_1Hzflares} top panel). During
other observations, the 1 Hz QPO is much broader and it wings blend
with the first overtone (Fig.~\ref{fig:1808_2002_1Hzflares} middle two
panels). In addition to the 1 Hz QPO, QPOs around 30--40 Hz are
sometimes seen (see also \cite{vanstraaten2004}). It is unclear if
this 30--40 Hz QPO is related to the low-frequency QPOs discussed
above or if it is due to a different mechanism.  During certain
observations the 1 Hz QPO becomes very broad, turning into a
band-limited noise component (Fig.~\ref{fig:1808_2002_1Hzflares}
bottom panel).  A detailed analysis of the 1 Hz flaring phenomenon is
in progress.

\subsubsection{Observations at other wavelengths}

Rupen et al. \cite{rupen2002} detected the source at radio wavelengths
during the 2002 outburst. On October 16--17, 2002, they found a
0.3--0.44-mJy source at 8.5 GHz. Monard \cite{monard2002} reported
that on October 16 the optical counterpart was detectable at
magnitudes similar to those observed at the peak of the 1998 outburst.

\section{XTE J1751--305}

The second accretion-driven millisecond pulsar (XTE J1751--305) was
discovered on April 3, 2002 \cite{markwardt2002}. Its spin frequency
is 435 Hz (Fig.~\ref{fig:pulsations}) and the neutron star is in a
very small binary with an orbital period of 42 minutes. The timing
analysis of the pulsations gave a minimum mass for the companion of
0.013 \mzon~and a pulse-frequency derivative of $<3 \times 10^{-13}$
Hz s$^{-1}$. Assuming that the mass transfer was driven by
gravitational radiation, the distance toward the source could be
constrained to be $>$7 kpc with a mass for the companion of
0.013--0.035 \mzon, which suggests a heated helium dwarf
\cite{markwardt2002}. {\itshape Chandra} also observed the source,
resulting in an arc-second position \cite{markwardt2002}.

\begin{figure}
\includegraphics[width=7.5cm]{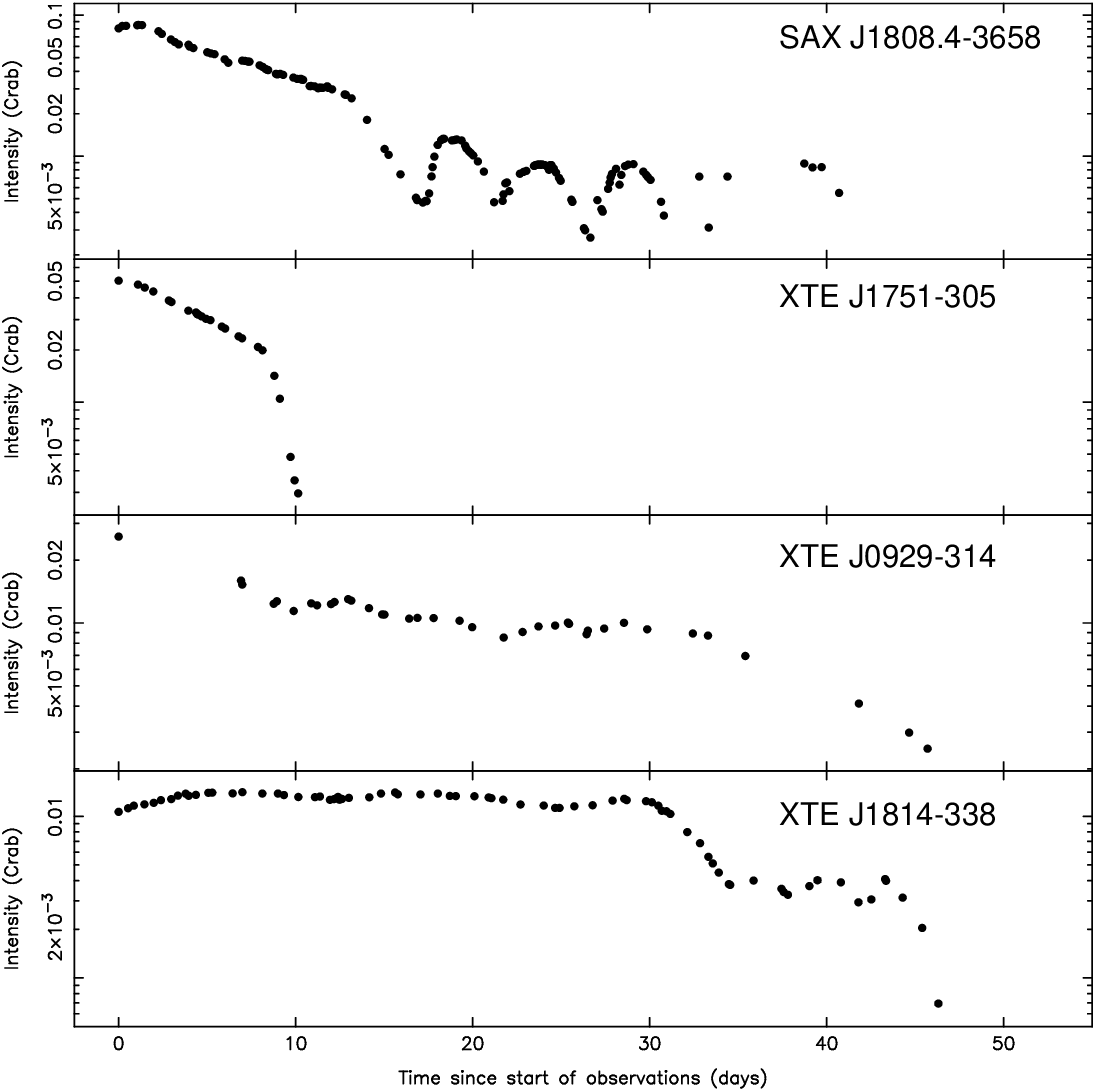}
\caption{
The {\it RXTE}/PCA light curve of SAX J1808.4--3658, XTE J1751--305,
XTE J0929--314, and XTE J1814--338. The data for SAX J1808.4--3658 was
obtained during its 2002 outburst. The data were taken from van
Straaten et al. \cite{vanstraaten2004} \label{fig:lightcurves}}
\end{figure}

The source reached a peak luminosity of $>$2$\times 10^{37}$ \Lunit,
an order of magnitude brighter than the peak luminosity of SAX
J1808.4--3658. However, the outburst was very short with an e-folding
time of only $\sim$7 days (compared to $\sim14$ days for SAX
J1808.4--3658; Fig.~\ref{fig:lightcurves}) resulting in a low outburst
fluence of only $\sim2.5 \times 10^{-3}$ ergs cm$^{-2}$
\cite{markwardt2002}. A potential re-flare was seen two weeks after
the end of the outburst during which also a type-I X-ray burst was
seen. Analysis of the burst indicated that the burst did not come from
XTE J1751--305 but it likely originated from the bright X-ray
transient in Terzan 6 \cite{intzandterzan}. It was also determined
that the transient in Terzan 6 could not have produced the re-flare
\cite{intzandterzan} suggesting that this re-flare could still have
come from XTE J1751--305. However, van Straaten et
al. \cite{vanstraaten2004} suggested (based on a X-ray color study
using {\it RXTE}/PCA) that this re-flare is caused by one of the
background sources and not by XTE J1751--305.  Van Straaten et
al. \cite{vanstraaten2004} also investigated the aperiodic timing
properties of the source (an example power spectrum is shown in
Fig.~\ref{fig:broad-band}) and the correlations between the
characteristic frequencies of the observed power-spectral
components. The frequency correlations were similar to those of the
non-pulsating neutron-star LMXBs. In contrast with the results
obtained for SAX J1808.4--3658 (see above), no frequency shift was
required for XTE J1751--305 to make the frequency correlations
consistent with those of the non-pulsating sources. Using these
correlations, van Straaten et al. \cite{vanstraaten2004} suggested
that the highest-frequency noise components in XTE J1751--305 are
likely due to the same mechanisms as the kHz QPOs.  They also
investigated the correlations between the characteristic frequencies
and the X-ray colors of the source and concluded that it did not
behave like an atoll source.

A previous outburst in June 1998 was detected using archival {\it
RXTE}/ASM data \cite{markwardt2002}, suggesting a tentative recurrence
time of $\sim$3.8 years.  Miller et al. \cite{miller2002} reported on
high spectral resolution data of the source obtained with {\itshape
XMM-Newton}. They only detected a continuum spectrum dominated by a
hard power-law shaped component (power-law index of $\sim$$1.44$) with
a 17 \% contribution to the 0.5--10 keV flux by a soft thermal
(black-body) component with temperature of $\sim$$1$ keV. Searches for
the optical and near-infrared counterparts were performed but no
counterparts were found
\cite{jonker2003}, likely due to the high reddening toward the
source. These non-detections did not constrain any models for the
accretion disk or possible donor stars.

\section{XTE J0929--314}

\begin{figure}
\includegraphics[width=7.5cm]{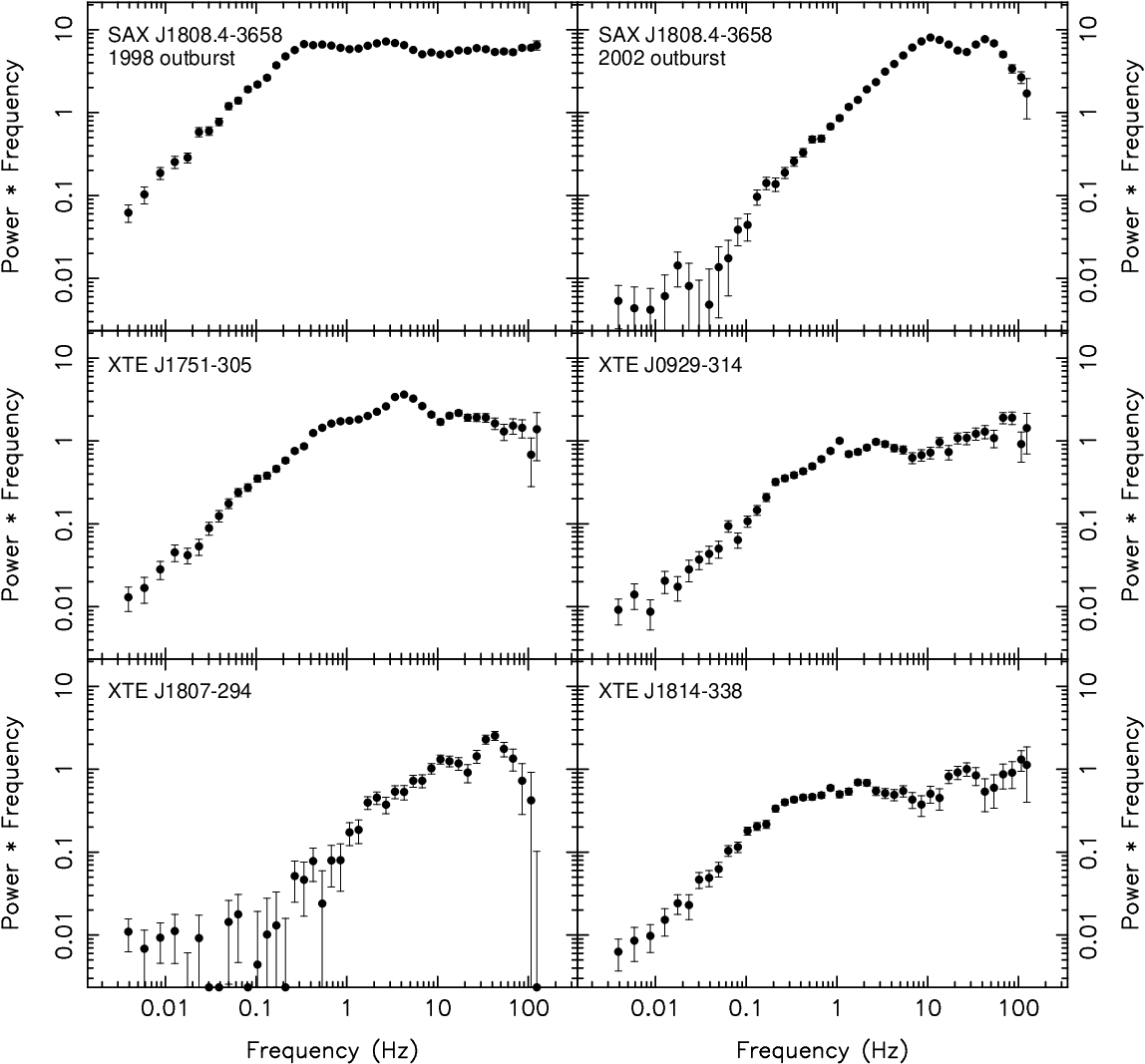}
\caption{
Examples of the aperiodic timing features seen in the five millisecond
pulsars (only for frequencies below $\sim$200 Hz). For SAX
J1808.4--3658 an example is shown for its 1998 and 2002 outbursts.
\label{fig:broad-band}}
\end{figure}

The third accretion-driven millisecond X-ray pulsar XTE J0929--314 was
already detected with the {\itshape RXTE}/ASM on April 13, 2002
\cite{remillard2002}, but not until May 2 (when
observations of the source were made using the {\it RXTE}/PCA) was it
found to be harboring a millisecond pulsar with a pulsation frequency
of 185 Hz (\cite{remillardetal2002};
Fig.~\ref{fig:pulsations}). Galloway et al. \cite{galloway2002}
reported on the detection of the 44-min orbital period of the
system. A minimum mass of 0.008 \mzon~was obtained for the companion
star and a pulse-frequency derivative of ($-9.2\pm0.4)\times 10^{-14}$
Hz s$^{-1}$. Galloway et al. \cite{galloway2002} suggested that this
spin down torque may arise from magnetic coupling to the accretion
disk, a magneto-hydrodynamic wind, or gravitational radiation from the
rapidly spinning neutron star.  Assuming gravitational radiation as
the driving force behind the mass transfer, Galloway et
al. \cite{galloway2002} found a lower limit to the distance of 6
kpc. They also reported on the detection of a QPO at 1 Hz
(Fig.~\ref{fig:broad-band}).  Full details of this QPO and the other
aperiodic power-spectral components are presented by van Straaten et
al. \cite{vanstraaten2004}. Just as they found for SAX J1808.4--3658,
the frequency correlations for XTE J0929--314 were similar to those
observed for the non-pulsating sources but with an offset in the
frequencies of the highest-frequency components. These correlations
allowed van Straaten et al. \cite{vanstraaten2004} to identify those
components as related to the kHz QPOs. Studying the correlated
spectral and timing variability, they concluded that the behavior of
XTE J0929--314 was consistent with that of an atoll source.

Juett et al. \cite{juett2003} obtained high resolution spectral data
using the {\itshape Chandra} gratings. Again the spectrum is well
fitted by a power-law plus a black body component, with a power-law
index of 1.55 and a temperature of 0.65 keV. Similar to XTE
J1751--305, no emission or absorption features were found. No orbital
modulation of the X-ray flux was found implying an upper limit on the
inclination of 85\degr. Greenhill et al. \cite{greenhill2002} reported
the discovery of the optical counterpart of the system. Castro-Tirado
et al. \cite{ct2002} obtained optical spectra of the source on May
6--8 in the range 350--800 nm and found emission lines from the C III
- N III blend and H-alpha, which were superposed on a blue
continuum. These optical properties are typical of X-ray transients
during outburst.  Rupen et al. \cite{rupen2002_0929} discovered the
radio counterpart of the source using the VLA with a 4.86 GHz flux of
0.3--0.4 mJy.

\section{XTE J1807--294}

The fourth millisecond X-ray pulsar XTE J1807--294 with a frequency of
191 Hz, was discovered on February 21, 2003
(\cite{markwardt2003_1807}; Fig.~\ref{fig:pulsations}). The peak flux
was only 58 mCrab (2--10 keV). The orbital period was determined
\cite{markwardt2003_atel} to be $\sim$$40$ minutes making it the
shortest period of all accretion-driven millisecond pulsars
known. Using a {\itshape Chandra} observation, Markwardt et
al. \cite{markwardt2003_atel} reported the best known position of the
source. In the {\it RXTE}/PCA data, kHz QPOs have been detected for
this system and the results obtained from a full analysis of those
data will be reported elsewhere \cite{markwardt2004_prep}. An example
of the power-spectral components at frequencies below 200 Hz is show
in Figure~\ref{fig:broad-band}. Campana et al. \cite{campana2003_1807}
reported on a {\itshape XMM-Newton} observation of this source taken
on March 22, 2003. Assuming a distance of 8 kpc, the 0.5--10 keV
luminosity during that observation was $2\times10^{36}$ \Lunit. They
could detect the pulsations during this observation with a pulsed
fraction of 5.8\% in the 0.3--10 keV band (increasing with energy) and
a nearly sinusoidal pulse profile (see also \cite{kk2003}).  The
spectral data are well fit by a continuum model, assumed to be an
absorbed Comptonisation model plus a soft component. The latter
component only contributed 13\% to the flux. Again no emission or
absorption lines were found.  No detections of the counterparts of the
system at other wavelengths have been reported so far.

\section{XTE J1814--338}

The fifth system (XTE J1814--338) was discovered on June 5, 2003 and
has a pulse frequency of 314 Hz (\cite{markwardt2003_1814};
Fig.~\ref{fig:pulsations}), with an orbital period of 4.3 hr and a
minimum companion mass of 0.15 \mzon
\cite{markwardt2003_1814_atel}. This 4.3 hr orbital period makes it
the widest binary system among the accretion-driven millisecond
pulsars and also the one most similar to the general population of the
low-luminosity neutron-star LMXBs (the atoll sources).  Many type-I
X-ray bursts with burst oscillations were found with a frequency
consistent with the neutron star spin frequency
\cite{markwardt2003_1814_atel,strohmayer2003}. A distance of $\sim$8
kpc was obtained from a burst which likely reached the Eddington
luminosity. The burst oscillations are strongly frequency and phase
locked to the persistent pulsations (as was also seen for SAX
J1808.4--3658; \cite{chakrabarty2003}) and two bursts exhibited
evidence for a frequency decrease of a few tenths of a Hz during the
onset of the burst, suggesting a spin down. Strohmayer et
al. \cite{strohmayer2003} also reported on the detection of the first
harmonic of the burst oscillations: the first time that this has been
found for any burst-oscillation source. This harmonic could arise from
two hot-spots on the surface, but they suggested that if the burst
oscillations arise from a single bright region, the strength of the
harmonic would suggest that the burst emission is beamed (possibly due
to a stronger magnetic field strength than in non-pulsating LMXBs).

Wijnands \& Homan \cite{wh2003} analyzed the {\itshape RXTE}/PCA data
of the source obtained between June 8 and 11, 2003. The overall shape
of the 3-60 keV power spectrum is dominated by a strong broad
band-limited noise component (Fig.~\ref{fig:broad-band}), which could
be fitted by a broken power-law model with a broad bump superimposed
on it at frequencies above the break frequency. Van Straaten et
al. \cite{vanstraaten2004} performed an in-dept analysis of all
publicly available {\it RXTE}/PCA data of the source to study the
power-spectral components and the correlations between their
characteristic frequencies. Using those correlations and by comparing
them to other sources, they could identify several components as
related to the kHz QPOs. They also found that the frequency
correlations were identical to the non-pulsating sources with no need
for a frequency shift. This is similar to what they found for XTE
J1751--305 but different from SAX J1808.4--3658 and XTE
J0929--314. The reason(s) for this difference between accreting
millisecond pulsars is not know (see
\cite{vanstraaten2004} for a discussion). From the correlations
between the spectral and timing variability it was also conclude that
the behavior of XTE J1814--338 was consistent with that of an atoll
source (\cite{vanstraaten2004}; see also \cite{wh2003}).

Wijnands \& Reynolds \cite{wr2003} reported that the position of XTE
J1814--338 was consistent with the {\itshape EXOSAT} slew source EXMS
B1810--337 which was detected on September 2, 1984. If XTE J1814--338
can indeed be identified with EXMS B1810--337, then its recurrence
time can be inferred to be less than 19 years but more than 5 years
(the time since the {\itshape RXTE}/PCA bulge scan observations
started in February 1999), unless the recurrence time of the source
varies significantly. Krauss et al. \cite{krauss2003} reported the
best position of the source as obtained using {\itshape Chandra} and
on the detection of the likely optical counterpart (with magnitudes of
B = 17.3 and R = 18.8 on June 6). Steeghs \cite{steeghs} performed
optical spectroscopy of this possible counterpart and reported
prominent hydrogen and helium emission lines, confirming the
connection between the optical source and XTE J1814--338.

\section{Concluding remarks}

From the current review it is clear that {\it RXTE} is vital to the
discovery and study of accretion-driven millisecond X-ray
pulsars. Thanks to {\it RXTE} we know about the existence of five such
systems. The detailed studies performed with {\it RXTE} for those
systems have yielded break-throughs in our understanding of kHz QPOs
and burst oscillations. Furthermore, three of these accreting pulsars
are in ultrashort binaries which will constrain evolutionary paths for
this type of system (e.g., see \cite{np2003}). However, it is also
clear that the five systems do not form a homogeneous group; their
pulsation frequencies span the range between 185 Hz and 435 Hz
(Fig.~\ref{fig:pulsations}), their orbital periods between 40 minutes
and 4.3 hrs, and their X-ray light curves are very different
(Fig.~\ref{fig:lightcurves}). More well studied outbursts of the
currently known systems are needed as well as discoveries of
additional systems. At the moment, only {\it RXTE} is capable of
performing the necessary timing observations. After {\it RXTE} the
need for an instrument with at least similar or better capabilities is
highly desirable for our understanding of accretion-driven millisecond
pulsars and their connection with the non-pulsating neutron-star
LMXBs.

For lack of space this review has focused on observational findings on
accretion-driven millisecond pulsars during outbursts. For information
on their quiescent states and on theoretical progress, I refer to
\cite{wijnands2003_review}.

\begin{theacknowledgments}
I thank Craig Markwardt for kindly providing the {\it RXTE}/PCA data
on XTE J1807--294 which I used to make Figures~\ref{fig:pulsations}
and \ref{fig:broad-band}. I also thank Steve van Straaten for help in
making Figures~\ref{fig:1808_lc_2002}, \ref{fig:1808_2002_1Hzflares},
and \ref{fig:lightcurves}.
\end{theacknowledgments}




\begin{thebibliography}{}

\bibitem{bvdh1991}Bhattacharya, D. \& van den Heuvel, E.P.J. 
1991, Ph.R. 203, 1

\bibitem{campana2003_1807}Campana, S. et al. 2003, ApJ, 594, L39

\bibitem{ct2002}Castro-Tirado, A.J. et al. 2002, IAUC 7895

\bibitem{cm1998}Chakrabarty, D. \& Morgan, E. H. 1998, Nature, 394, 346

\bibitem{chakrabarty2003}Chakrabarty, D. et al. 2003, Nature, 424, 42

\bibitem{cui1998}Cui, W. et al. 1998, ApJ, 504, L27

\bibitem{ford1999}Ford, E.C. 1999, ApJ, 519, L73

\bibitem{gaensler1999}Gaensler, B.M. et al. 1999, ApJ, 522, L117

\bibitem{galloway2002}Galloway, D.K. et al. 2002, ApJ, 576, L137

\bibitem{gierlinski2002}Gierlinski, M. et al. 2002, MNRAS, 331, 141

\bibitem{giles1998}Giles, A.B. et al. 1998, IAUC 6886

\bibitem{giles1999}Giles, A.B. et al. 1999, MNRAS, 304, 47

\bibitem{gilfanov1998}Gilfanov, M. et al. 1998, A\&A, 338, L83

\bibitem{greenhill2002}Greenhill, J.G. et al. 2002, IAUC 7889

\bibitem{hs1998}Heindl, W.A. \& Smith, D.M. 1998, ApJ, 506, L35

\bibitem{intzand1998}In 't Zand, J.J.M. et al. 1998, A\&A, 331, L25

\bibitem{intzand2001}In 't Zand, J.J.M. et al. 2001, A\&A, 372, 916

\bibitem{intzandterzan}In 't Zand, J.J.M. et al. 2003, A\&A, 409, 659

\bibitem{jonker2003}Jonker, P.G. et al. 2003, MNRAS, 344, 201

\bibitem{juett2003}Juett, A.M. et al. 2003, ApJ, 587, 754

\bibitem{kazarovets2000}Kazarovets, E.V. et al. 2000, IBVS, 4870

\bibitem{kk2003}Kirsch, M.G.F. \& Kendziorra, E. 2003, ATEL 148

\bibitem{kluzniak2003}Klu\'zniak, W. et al. 2003, astro-ph/0308035

\bibitem{krauss2003}Krauss, M.I. et al. 2003, IAUC 8154

\bibitem{lm2003}Lamb, F.K. \& Miller, M.C. 2003, ApJ Letters, 
submitted (astro-ph/0308179)

\bibitem{markwardt2002}Markwardt, C.B. et al. 2002, ApJ, 575, L21

\bibitem{markwardt2002_1808}Markwardt, C.B. et al. 2002, IAUC 7993

\bibitem{markwardt2003_1807}Markwardt, C.B. et al. 2003, IAUC 8080

\bibitem{markwardt2003_1814}Markwardt, C.B. et al. 2003, IAUC 8144

\bibitem{markwardt2003_atel}Markwardt, C.B. et al. 2003, ATEL 127

\bibitem{markwardt2003_1814_atel}Markwardt, C.B. et al. 2003, ATEL 164

\bibitem{markwardt2004_prep}Markwardt, C.B. et al. 2004, ApJ in preparation

\bibitem{marshall1998}Marshall, F.E. 1998, IAUC 6876

\bibitem{miller2002}Miller, J.M. et al. 2003, ApJ, 583, L99

\bibitem{monard2002}Monard, B. 2002, VSNet alert 7550

\bibitem{morgan2004}Morgan, E.H. et al. 2004, ApJ in preparation

\bibitem{np2003}Nelson, L. A. \& Rappaport, S. 2003, ApJ, 598, 431

\bibitem{poutanengierlinski2003}Poutanen, J. \& Gierlinski, M. 2003, 
MNRAS 343, 1301

\bibitem{remillard2002}Remillard, R.A. 2002, IAUC 7888

\bibitem{remillardetal2002}Remillard, R.A. et al. 2002, IAUC 7893

\bibitem{revnivtsev2003}Revnivtsev, M.G. 2003, AstL, 29, 383

\bibitem{roche1998}Roche, P. et al. 1998, IAUC 6885

\bibitem{rupen2002_0929}Rupen, M.P. et al. 2002, IAUC 7893

\bibitem{rupen2002}Rupen, M. et al. 2002, IAUC 7997

\bibitem{steeghs}Steeghs, D. 2003, IAUC 8155

\bibitem{strohmayer2003}Strohmayer, T.E. et al. 2003, 596, L67

\bibitem{vdk2000}Van der Klis, M. 2000, ARA\&A, 38, 717

\bibitem{vdk2000_1808}Van der Klis, M. et al. 2000, IAUC, 7358

\bibitem{vanstraaten2003}Van Straaten, S. et al. 2003,  in 'The Restless 
High-Energy Universe', 5--8 May 2003, Amsterdam, the Netherlands
(eds. E.P.J. van den Heuvel, J.J.M. in 't Zand, \& R.A.M.J. Wijers)
(astro-ph/0309345)

\bibitem{vanstraaten2004}Van Straaten, S., et al. 2004, ApJ, submitted

\bibitem{v1994}Vaughan, B.A. et al. 1994, ApJ, 435, 362

\bibitem{wh2000}Wachter, S. \& Hoard, D.W. 2000, IAUC 7363

\bibitem{wachter2000}Wachter, S. et al. 2000, HEAD 32, 24.15

\bibitem{wang2001}Wang, Z. et al. 2001, ApJ, 563, L61

\bibitem{wijnands2001_flaring}Wijnands, R. 2001, Adv. Space Res. 28, 469

\bibitem{wijnands2003}Wijnands, R. 2003, ApJ, 588, 425

\bibitem{wijnands2003_review}Wijnands, R. 2003, in 'The Restless 
High-Energy Universe', 5--8 May 2003, Amsterdam, the Netherlands
(eds. E.P.J. van den Heuvel, J.J.M. in 't Zand, \& R.A.M.J. Wijers)
(astro-ph/0309347)

\bibitem{wvdk1998}Wijnands, R. \& van der Klis, M. 1998, Nature, 
394, 344

\bibitem{wvdk1998_bbn}Wijnands, R. \& van der Klis, M. 1998, ApJ, 507, L63

\bibitem{wh2003}Wijnands, R. \& Homan, J. 2003, ATEL  165

\bibitem{wr2003}Wijnands, R. \& Reynolds, A. 2003, ATEL 166

\bibitem{wijnands2001}Wijnands, R. et al. 2001, ApJ, 560, 892

\bibitem{wijnands2002}Wijnands, R. et al. 2002, ApJ, 571, 429

\bibitem{wijnands2003_nature}Wijnands, R. et al.2003, Nature, 424, 44




\IfFileExists{\jobname.bbl}{}
 {\typeout{}
  \typeout{******************************************}
  \typeout{** Please run "bibtex \jobname" to optain}
  \typeout{** the bibliography and then re-run LaTeX}
  \typeout{** twice to fix the references!}
  \typeout{******************************************}
  \typeout{}
 }
\end{thebibliography}

\end{document}